\documentclass[fleqn,twoside,twocolumn,nofootinbib,showkeys,nosuperscriptaddress]{revtex4} % Specifies the document class %,unsortedaddress
\usepackage{color}
\usepackage{graphicx}
\def\autorcol#1{\def\@autorcol{#1}}
\begin{document}
\title[Central exclusive production at LHCb]
{Central exclusive production at LHCb}
\author{C. Van Hulse, on behalf of the LHCb Collaboration}
\affiliation{University College Dublin
  \\Belfield, Dublin 4, Ireland}
%\address{Belfield, Dublin 4, Ireland}
%\affiliation{Belfield, Dublin 4, Ireland}
\email{cvanhuls@mail.cern.ch}
%\razd{\seci}

\autorcol{C. Van Hulse, on behalf of the LHCb Collaboration}%

\setcounter{page}{1}%

\begin{abstract}
The LHCb collaboration has measured central exclusive production of $J/\psi$, $\psi(2S)$, and
$\Upsilon$ mesons as well as $J/\psi J/\psi$, $J/\psi\psi(2S)$, $\psi(2S)\psi(2S)$, and $\chi_c\chi_c$ meson pairs in proton-proton collisions.
The analyses of $\Upsilon$ and charmonium pairs are performed at the centre-of-mass energies of 7~TeV and 8~TeV,
and those of $J/\psi$ and $\psi(2S)$ are done at 7~TeV and 13~TeV.
The analysis at 13~TeV involves the use of new shower counters. These allow a reduction in the background by
vetoing events with activity in an extended region in rapidity.
The measurements of central exclusive production at LHCb are sensitive to gluon distributions for Bjorken-$x$ values down to $2\times10^{-6}$ (at 13~TeV). An overview of the LHCb results is presented and compared to existing measurements of other experiments and theoretical calculations.
\end{abstract}


\maketitle

\section{Introduction}

\begin{figure*}% figure* for wide figure, [h] [!] to change the placement
\vskip-3mm
\begin{minipage}{0.48\textwidth}\centering
\includegraphics[width=0.9\textwidth]{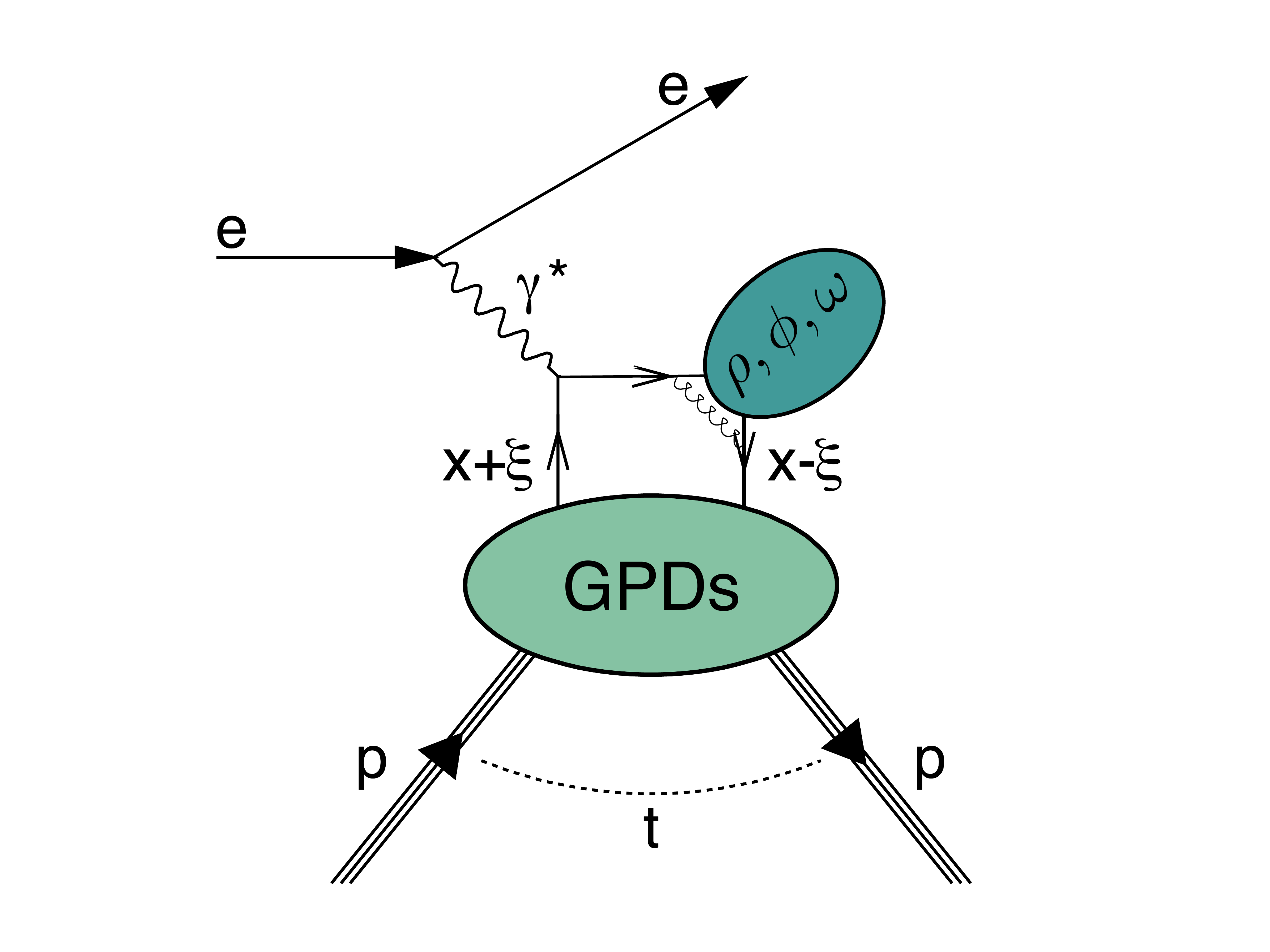}
\end{minipage}
\begin{minipage}{0.48\textwidth}\centering
\includegraphics[width=0.9\textwidth]{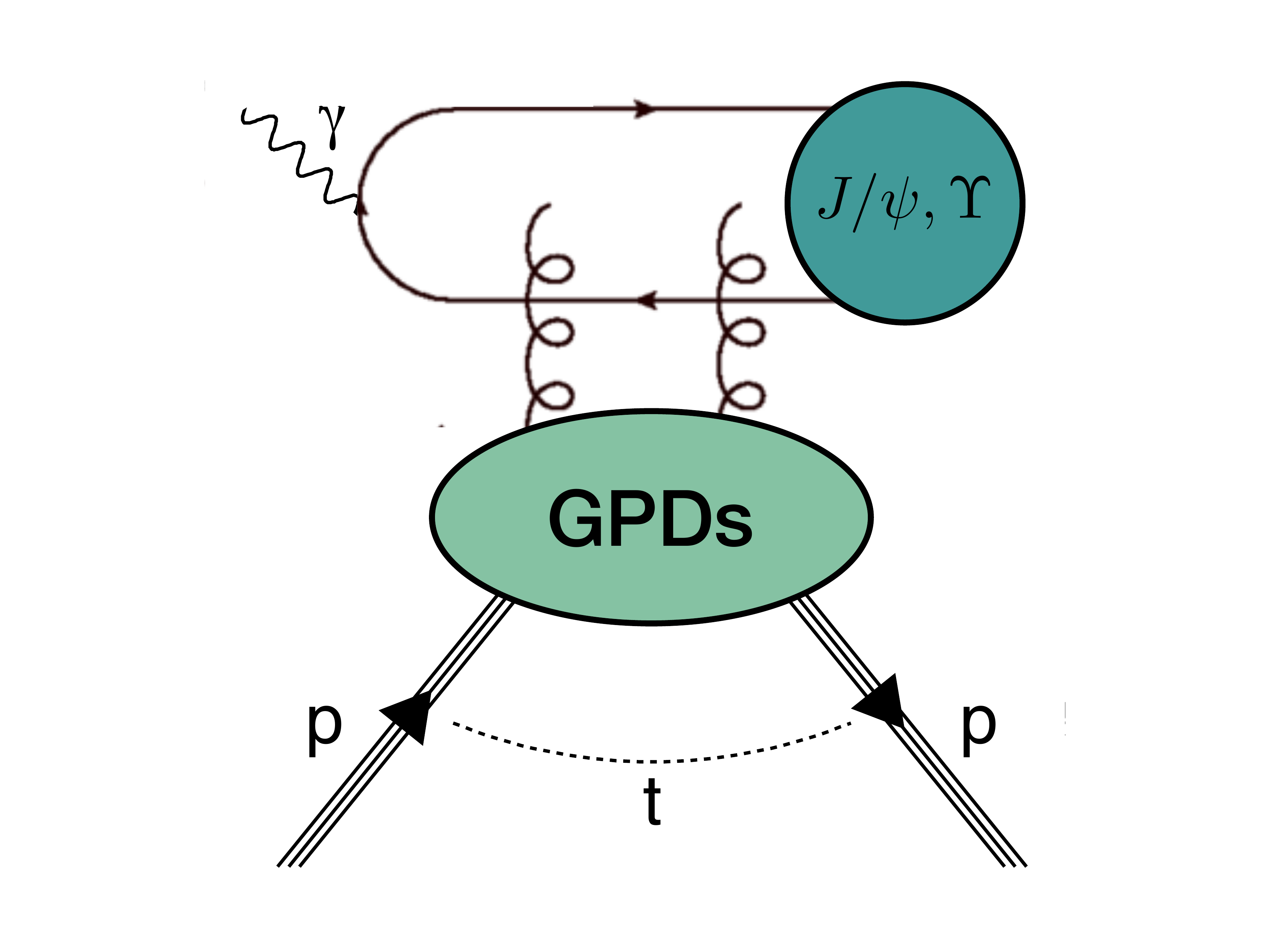}
\end{minipage}
\vskip-3mm\caption{Diagrams for exclusive production of vector mesons in deep-inelastic scattering (left) and in photoproduction (right). The 
figure on the left illustrates access to quark GPDs, while the figure on the right shows the diagram for gluon GPDs.}
\label{fig:hdbag}
\end{figure*}

The nucleon structure can be described in three dimensions in terms of the
probability to find quarks and gluons as a function of their transverse position inside the nucleon and 
their longitudinal momentum fraction with respect to the nucleon momentum~\cite{Bur1, Bur2}. The longitudinal direction coincides here with 
the direction of the probe used to investigate the nucleon.
The corresponding probability distributions are called impact-parameter-dependent parton distributions. They are Fourier transforms
of generalised parton distributions (GPDs) (see, e.g., Ref.~\cite{Diehl}). These GPDs do not have a probabilistic interpretation.
Instead, they represent probability amplitudes for a parton with longitudinal momentum fraction $x+\xi$ to be emitted from a nucleon
and a parton with longitudinal momentum fraction $x-\xi$ to be absorbed by the nucleon. The nucleon stays intact, but receives a 
four-momentum transfer squared equal to $-t$. This is represented in Fig.~\ref{fig:hdbag}, for quarks (left) and gluons (right). 

Generalised parton distributions are accessible in exclusive reactions, such as the exclusive production of photons or vector mesons, 
involving a hard scale. The hard scale is necessary in order to factorise the process into perturbatively calculable parts and 
non-perturbative parts, which are the GPDs and meson distribution amplitudes in the case of exclusive meson production~\cite{Col97, Ra97}.
Exclusive vector-meson production can be measured in deep-inelastic scattering, as illustrated in Fig.~\ref{fig:hdbag}, left.
The hard scale is provided here by the large virtuality, $Q^2=-q^2 \gg 1$~GeV$^2$, of the photon exchanged between the lepton and the nucleon. 
There exists a multitude of such measurements at fixed-target experiments~\cite{clas17a, clas17b, compass14, compass17,hallA11,her14,her15,her17} 
and at lepton-proton colliders, by the H1 and ZEUS collaborations~\cite{h105,h106,zeus04,zeus07}.
The former series of measurements are mainly sensitive to larger values of Bjorken-$x$, $x_B$, with $\xi \approx x_{B}/(2-x_B)$, and thus to quark GPDs, 
while the latter probe lower values of $x_B$ down to $10^{-4}$, where gluons dominate.

Alternatively, it is possible to use a (quasi-)real photon ($Q^2\approx 0$ GeV$^2$) to investigate the nucleon, provided that the particle created in the 
final state has a large mass component. In the case of exclusive vector-meson production, such as $J/\psi$ or $\Upsilon$ production, 
the large scale is then provided by the large mass of the meson valence quarks (charm or bottom quarks).
The vector mesons originate, as illustrated in  Fig.~\ref{fig:hdbag}, right, from the splitting of the real photon into a quark-antiquark pair ($c\bar{c}$ or $b\bar{b}$).
This pair interacts with a nucleon through the exchange of two gluons, and as a result a vector meson is formed in the final state. 
%This process is 
%Fig.~\ref{fig:hdbag}, right. The photon splits into a quark-antiquark pair. This subsequently interacts with the nucleon through the exchange
%of two gluons, and as a result a $J/\psi$ or $\Upsilon$ is produced. 
Quasi-real photoproduction of vector mesons has been measured in electron-proton collisions by the H1 and ZEUS 
experiments~\cite{h1ph02,h1ph06,h1ph13,zeus04,zeusph09}.
These measurements probe a photon-nucleon center-of-mass energy ranging from 30~GeV to 300~GeV.

It is also possible to study photoproduction in ultra-peripheral collisions of protons and ions. 
In such reactions, the beam particles interact at a large enough distance from each other (in practice more than the sum of their respective 
nuclear radii) so that they interact through the exchange of colour-neutral objects. 
The flux of photons emitted by a beam particle is proportional
to the square of its atomic charge, and hence photon emission by heavy ions is greater than for protons. 
There exist measurements of exclusively produced vector mesons in gold-gold collisions by the PHENIX experiment~\cite{phen09}, 
in proton-antiproton collisions by the CDF experiment~\cite{cdf09}, 
in lead-lead and proton-lead collisions by the ALICE experiment~\cite{alice13,alice13b,alice14} and in proton-proton and lead-lead  collisions by the 
LHCb experiment~\cite{lhcb13,lhcb14,lhcb15,lhcb18,lhcblead}. The covered photon-nucleon center-of-mass energy ranges from 34~GeV, for the PHENIX experiment, to 1.5~TeV, for the measurements performed by the LHCb collaboration.
The very high energy available at the LHC offers the unique possibility to probe the GPDs down to Bjorken-$x$ values of the order of $10^{-6}$, i.e., two orders of magnitude lower than for the existing measurements in electron-proton collisions. At such low values of $x_B$, one might also be sensitive to 
saturation effects~\cite{arm14}. In addition, at such low $x_B$, the exclusive cross section can be approximated in terms of standard gluon 
parton distribution functions (PDFs)~\cite{Rys,jmrt,Jones,flett}. This cross section has a quadratic dependence on the gluon PDFs, 
and thus provides a higher sensitivity than inclusive measurements, where the dependence is only linear.

\section{LHCb measurements}

In central exclusive production of vector mesons in ultra-peripheral collisions, the proton (or ion) emitting the real photon is to a good approximation 
not altered from its original trajectory, while the proton interacting through the two gluons undergoes a small change in momentum, but remains close to the beam line. 
The vector meson, in turn, is produced in the central region. At LHCb, this vector meson is generally reconstructed through its decay into a $\mu^+\mu^-$ pair.
Hence, the experimental signature for exclusive vector-meson production 
are two oppositely charged muons in the LHCb detector, with large regions of rapidity, down to close to the beam line, devoid of particle activity.
There exists a different process with the same final state, but where the oppositely charged muons originate from the interaction of photons 
emitted by the respective beam particles. This process is called the Bethe--Heitler process.
This production mode of muons forms a continuum background to the exclusive production of vector mesons, 
and needs to be subtracted from the measured signal. Another source of background is the production of higher-mass vector mesons that decay 
into the vector meson under study without detection of the other decay products. Furthermore, the production of vector mesons 
where one or both of the interacting protons dissociate forms another background contribution.

The LHCb detector is a forward detector, covering a rapidity range between 2 and 5. The detector is fully instrumented for particle identification, 
and is capable of detecting particles with transverse momenta as low as 200~MeV. 
The LHCb experiment is not instrumented with detectors around the beam line for the detection of protons emerging intact from the interaction or 
for products from proton dissociation that remain close to the beam line. However, the LHCb experiment is well suited for the measurement of exclusive processes. Firstly, 
the average number of interactions per beam crossing at the LHCb interaction point ranges only from 1.1 to 1.5, depending 
on running conditions. Secondly, besides the coverage in rapidity from 2 to 5 by the fully instrumented LHCb detector, 
the LHCb vertex locator is capable of detecting charged-particle activity 
for rapidities between -3.5 and -1.5. Also, for Run 2 of the LHC data-taking period (2015 -- 2018), the LHCb experiment was additionally equipped with 
a series of five stations of scintillators, {\sc HeRSCheL}~\cite{herschel}, placed at -114~m to +114~m from the LHCb interaction point. This allowed for the
detection of particle showers in a rapidity range between -10 and -5, and between +5 and +10, and hence for a supplementary reduction of the contribution 
from background processes.

\begin{figure*}
\begin{minipage}{0.48\textwidth}\centering
\includegraphics[width=1.0\textwidth]{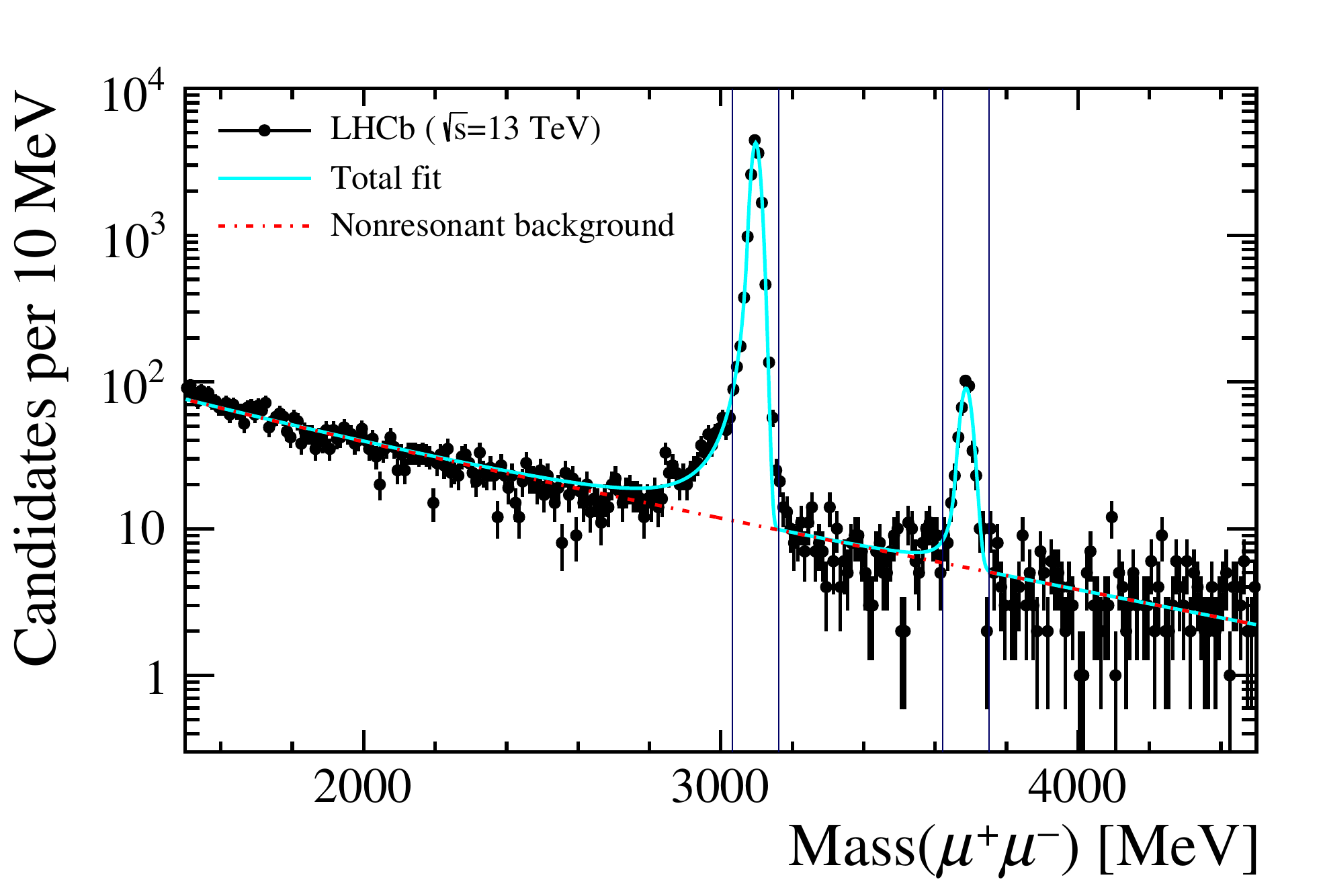}
\end{minipage}
\begin{minipage}{0.48\textwidth}\centering
\includegraphics[width=1.0\textwidth]{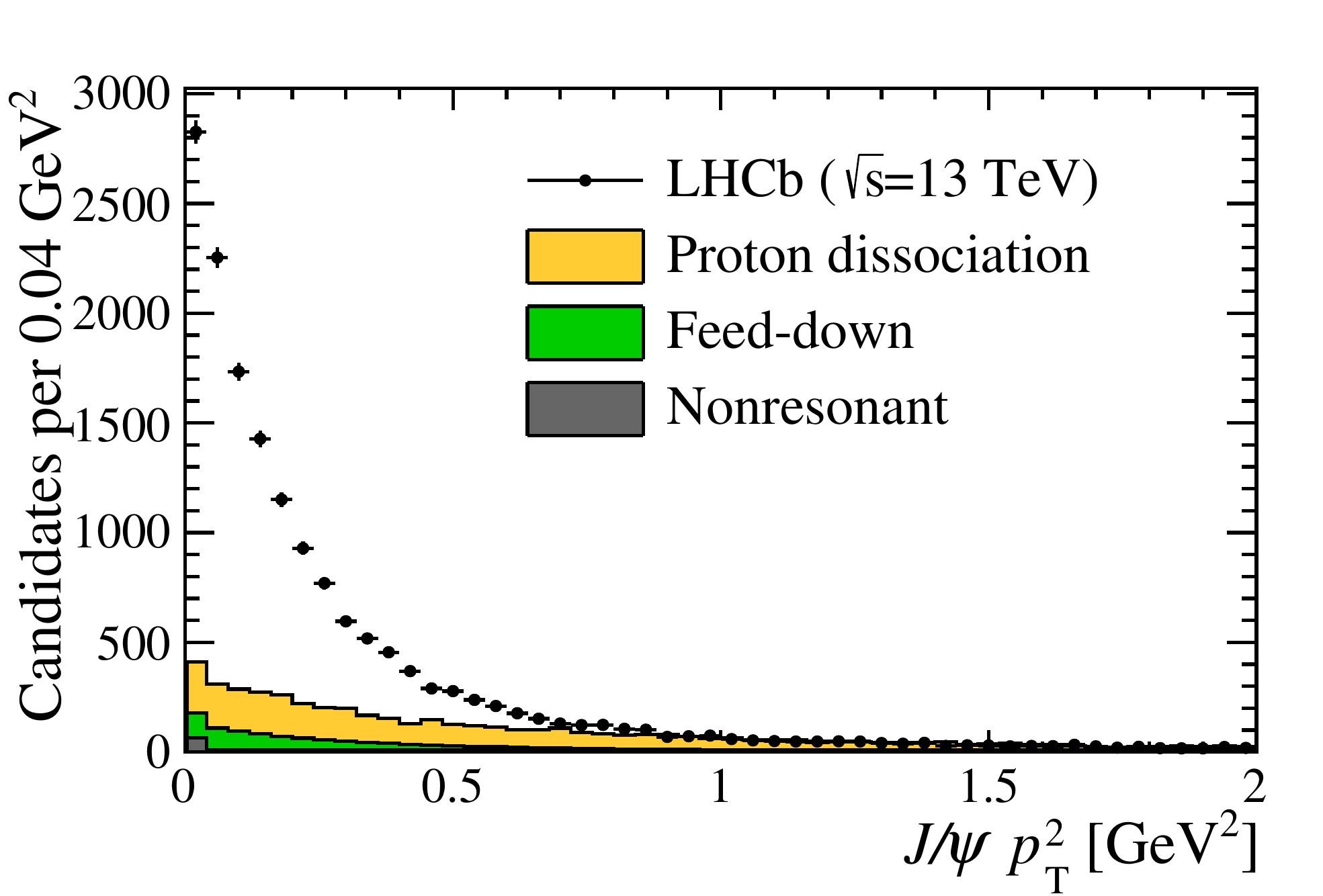}
\end{minipage}
\vskip-3mm\caption{Dimuon invariant-mass distribution (left) and dimuon squared-transverse-momentum distribution for muon pairs within the $J/\psi$ invariant mass region (right) for data collected at $\sqrt{s}=13$~TeV, 
and satisfying the selection requirement imposed by {\sc HeRSCheL}. Different background contributions are indicated in both figures, while
the vertical lines in the figure on the left indicate the selected range in invariant mass for the measurement of $J/\psi$ and $\psi(2S)$.}
\label{fig:dimumass}
\end{figure*}

Measurements of exclusive production of $J/\psi$ and $\psi(2S)$ mesons have been performed by the LHCb experiment using data collected
in proton-proton collisions at a center-of-mass energy $\sqrt{s}=7$~TeV~\cite{lhcb14}, and part of the data collected at $\sqrt{s}=13$~TeV~\cite{lhcb18}, 
amounting to an integrated luminosity of respectively $929\pm33$~pb$^{-1}$ and $204\pm8$~pb$^{-1}$. This data set allows to access $x_B$ down to $2\times10^{-6}$. Both the $J/\psi$ meson and the $\psi(2S)$ 
meson are reconstructed through their decay into muons, which are required to lie in the LHCb detector acceptance, between 2 and 4.5 in rapidity.
Furthermore, the transverse momentum squared of the dimuon pair, $p_{T}^2\approx -t$, needs to be below 0.8~GeV$^2$.
Finally, the absence of any other detector activity is required. 

\begin{figure}[!b]
\centering
\vskip-4mm\includegraphics[width=0.5\textwidth]{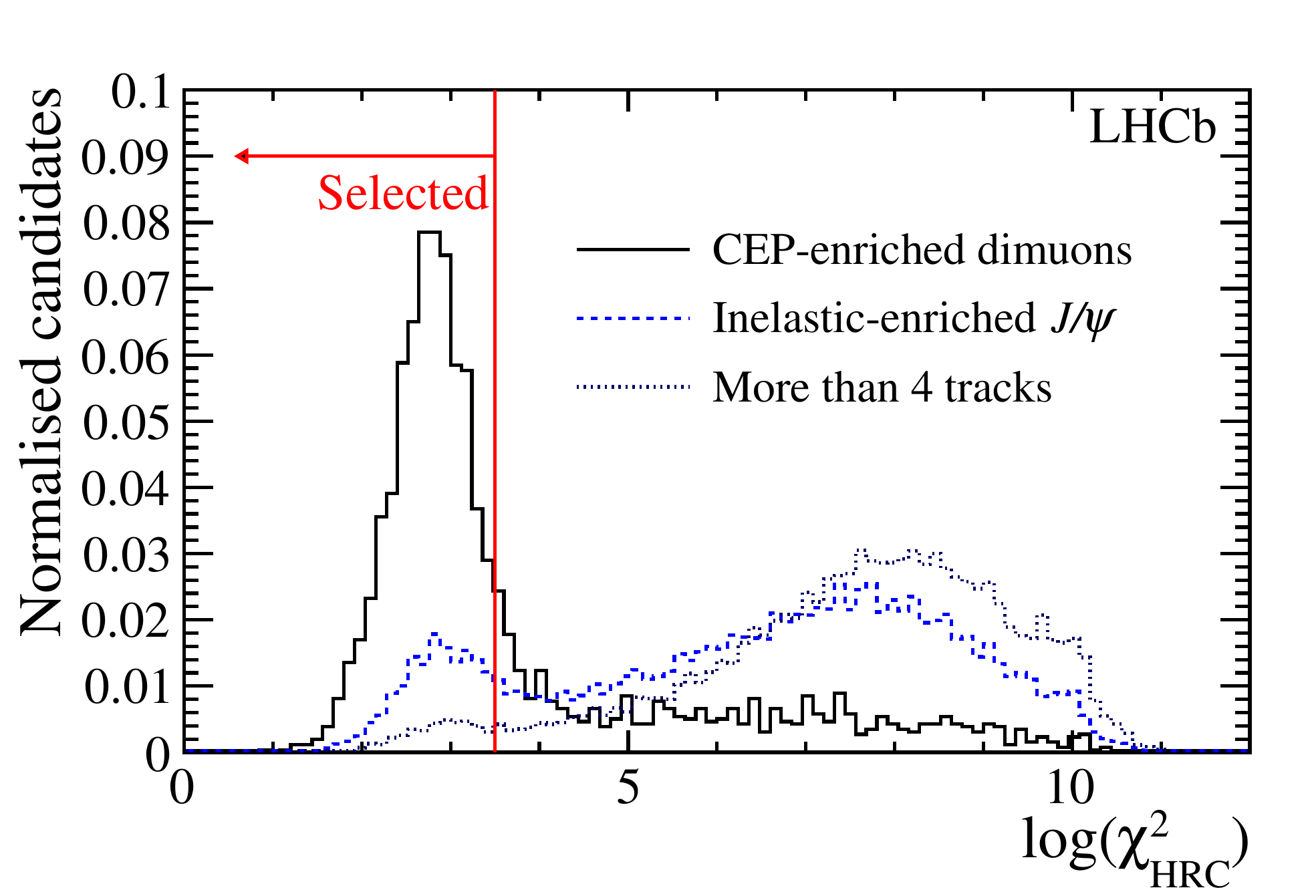}
\vskip-3mm\caption{The distribution, normalised to unit area, of the {\sc HeRSCheL} discriminating variable $\chi^2_{HRC}$. The continuous, black line corresponds to a sample highly enriched in exclusively produced muon pairs; the blue, dashed line represents the distribution for events enriched in inelastically produced $J/\psi$ mesons, while the purple, short-dashed line contains events with more than four tracks.}
\label{fig:herschel}
\end{figure}

In Fig.~\ref{fig:dimumass}, left, the dimuon invariant-mass distribution is shown, while in Fig.~\ref{fig:dimumass}, right, 
the squared-transverse-momentum distribution of the muon pair with invariant mass in the $J/\psi$ mass region is presented.
The three sources of background contamination to the $J/\psi$ signal are also shown.
The background contribution from the Bethe--Heitler process, labelled  nonresonant background,  is obtained 
from a fit to the dimuon mass distribution (see Fig.~\ref{fig:dimumass}, left). The background from feed-down from exclusive production of $\psi(2S)$ and $\chi_c$ mesons
%, where these particles decay into a $J/\psi$ and additional 
%particles remain undetected, 
is evaluated using $\psi(2S)$ and $\chi_c$ signals in experimental data and Monte-Carlo simulation. 
Finally, the contamination from events where at least one of the protons dissociates is evaluated for Run 1 through a 
fit of the dimuon transverse-momentum distribution, while for the data collected in Run 2, {\sc HeRSCheL} has been used. 
The discriminating power of {\sc HeRSCheL} is illustrated in Fig.~\ref{fig:herschel}. The figure represents distributions of a discriminating 
variable related to detector activity in {\sc HeRSCheL}. The continuous, black line is the distribution for a very pure sample of exclusively produced 
pairs of muons, while the other lines indicate samples enriched in nonexclusive events. From the figure, it is clear that for exclusive events, the discriminating 
variable is located at low values, whereas for nonexclusive events, the variable extends to higher values. For the selection of exclusive events in Run 2, 
only events below the value indicated by the red, vertical line are selected. This results in a signal purity of 76\% (73\%) for $J/\psi$ ($\psi(2S)$).
For data  collected in Run 1, the signal purity amounts to  62\% (52\%), where the contribution from proton-dissociative background is about twice as high.

\begin{figure*}
\begin{minipage}{0.48\textwidth}\centering
\includegraphics[width=0.9\textwidth]{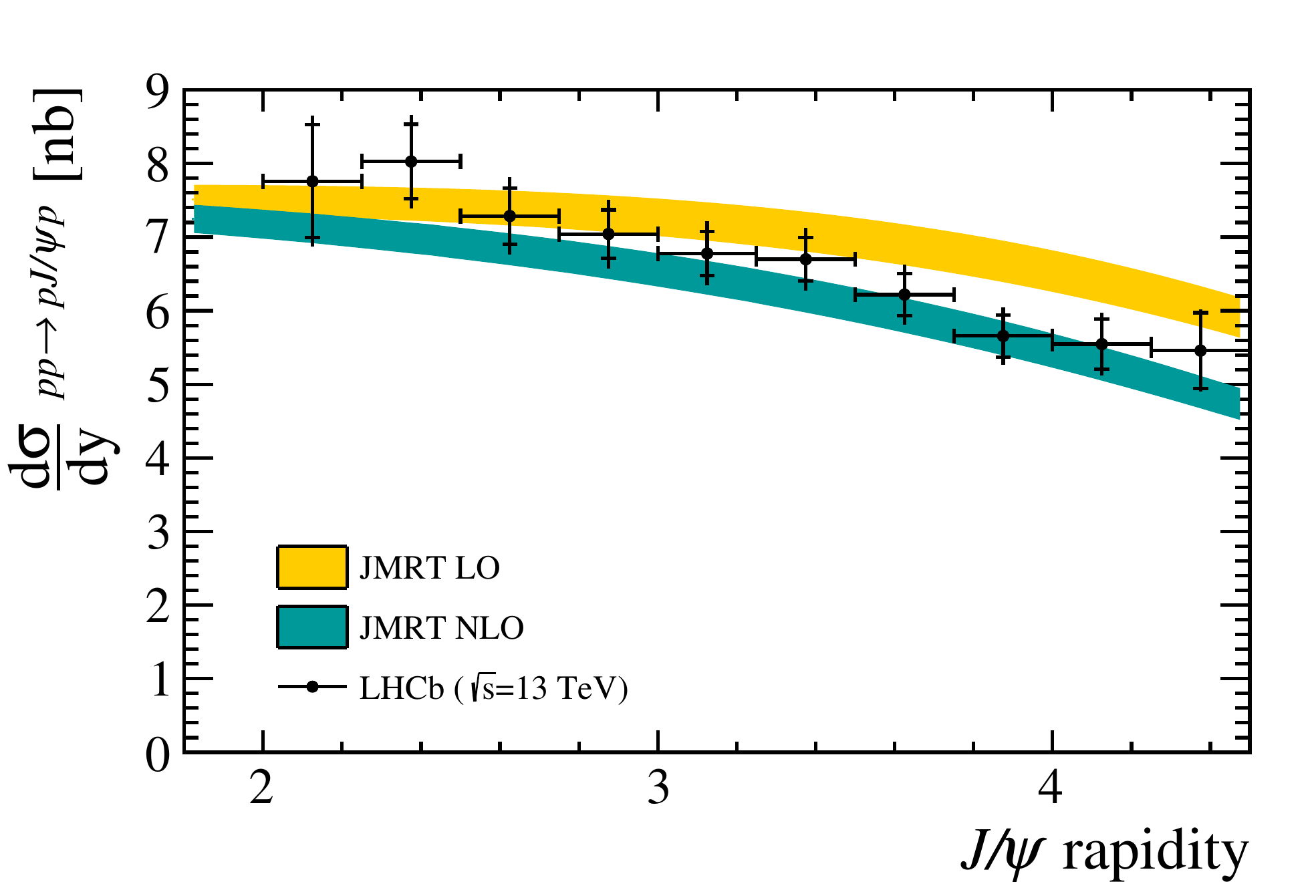}
\end{minipage}
\begin{minipage}{0.48\textwidth}\centering
\includegraphics[width=0.9\textwidth]{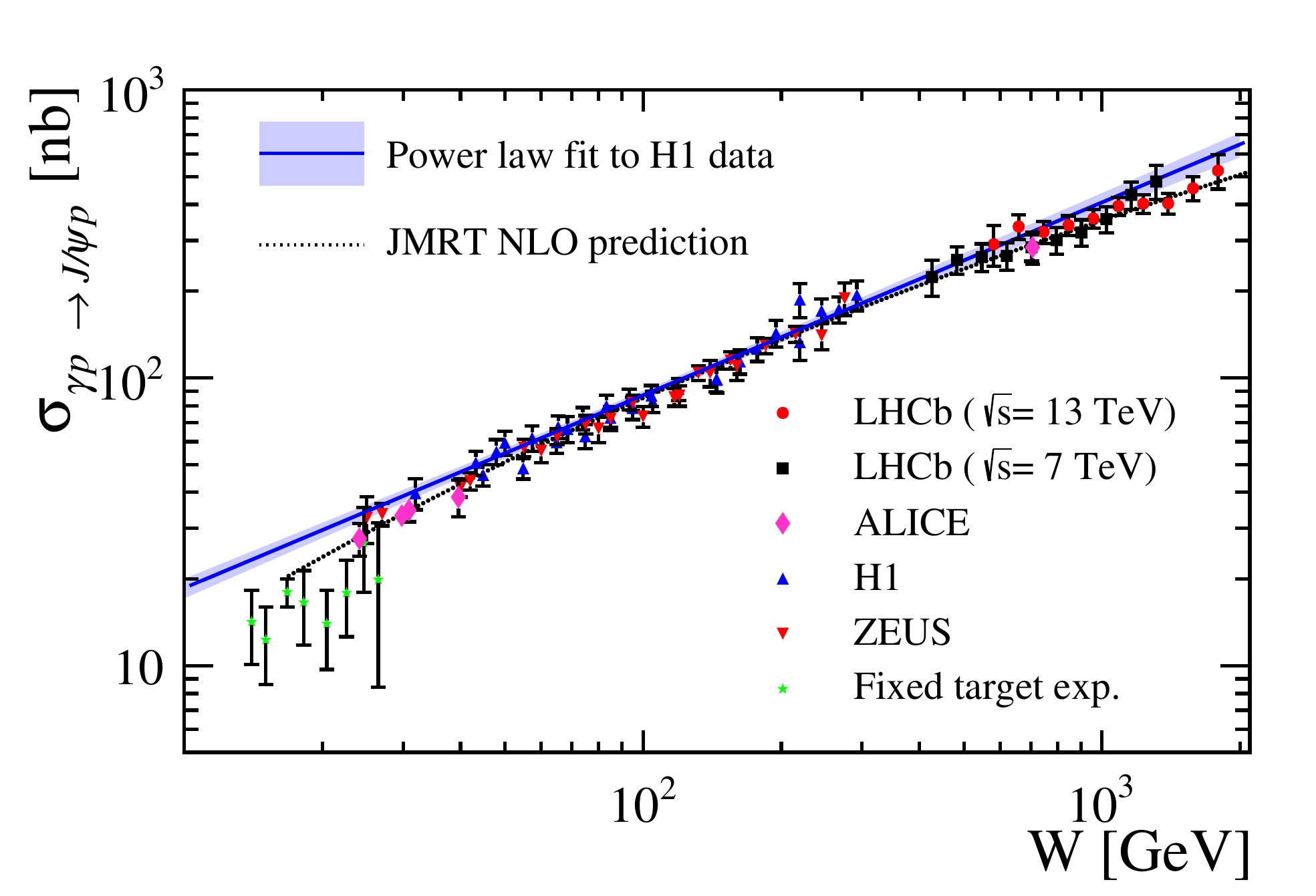}
\end{minipage}
\vskip-3mm\caption{Cross section differential in rapidity for exclusive $J/\psi$ production (left) and exclusive $J/\psi$ photoproduction cross section as a function of the photon-proton invariant mass (right). The leading-order (yellow band) and next-to-leading-order (green band and dotted line) 
JMRT calculations~\cite{jmrt} are also indicated.}
\label{fig:jpsi_cross}
\end{figure*}

The cross section differential in rapidity for exclusive production of $J/\psi$ in proton-proton collisions at $\sqrt{s}=13$~TeV 
is shown in Fig.~\ref{fig:jpsi_cross}, left. It is seen to decrease at larger values of rapidity. In addition to the experimental data points, 
theoretical predictions (JMRT)~\cite{jmrt}, which approximate the cross section in terms of standard gluon PDFs, are shown.
There are predictions at leading order in $\alpha_S$ (yellow band) and at next-to-leading order in $\alpha_S$ (green band). The leading-order 
predictions fail to describe the data at higher rapidities, while the next-to-leading order calculations are in reasonable agreement with the data.

The exclusive vector-meson production cross section in proton-proton collisions is related to the photoproduction cross section through 
\begin{eqnarray}
\sigma_{pp\rightarrow p\psi p} & = &  r(W_{+}) k_+\frac{dn}{dk_+}\sigma_{\gamma p \rightarrow \psi p} (W_+) \nonumber \\
 &+& r(W_{-}) k_-\frac{dn}{dk_-}\sigma_{\gamma p \rightarrow \psi p} (W_-), 
\label{eq:sigma}
\end{eqnarray}
where $r$ represents the gap survival factor, $k_{\pm}$ the photon energy, $dn/dk_{\pm}$ the photon flux, and $W^2_{\pm}=2 k_{\pm} \sqrt{s}$ the photon-proton invariant mass squared. The subscript $+$ ($-$) corresponds to the situation where the downstream-going (upstream-going) 
proton is the photon emitter.
As can be seen from Eq.~(\ref{eq:sigma}), the photoproduction cross section appears twice in the expression. The reason resides in the ambiguity on the identity of the proton emitting the real photon.
%, which could be either the upstream-going or downstream-going proton. 
Since the photoproduction cross section corresponding to the low-energy solution $W_-$ only contributes about one third of the time and  
it has been previously measured and parametrised by the H1 collaboration, this parametrisation is used to fix the low-energy photoproduction cross section, 
and extract the one at high photon-proton invariant mass. The resulting photoproduction cross section is presented in Fig.~\ref{fig:jpsi_cross}, right. The data points represented by the red circles
are the result at $\sqrt{s}=13$~TeV, while those indicated by the black squares are those at $\sqrt{s}=7$~TeV. Also measurements from the H1 collaboration, from which the parametrisation is taken, the ZEUS and ALICE collaborations as well as from fixed-target experiments~\cite{fix1,fix2,fix3} 
are presented. The different data sets are in good agreement with each other. The parametrisation from the H1 collaboration is indicated in the figure by the blue band. It is seen to 
describe the data well at intermediate values of $W$, but fails at lower and higher values. Also the next-to-leading order JMRT calculations are shown, 
as indicated by the dotted line. They are in good agreement with the data, describing it well also at low and high values of $W$.
Also for the proton-proton and photoproduction cross section of $\psi(2S)$ (not shown), the next-to-leading order predictions in $\alpha_S$ describe the data well,
whereas the leading-order predictions fail to describe the data.

\begin{figure*}
\begin{minipage}{0.48\textwidth}\centering
\includegraphics[width=0.89\textwidth]{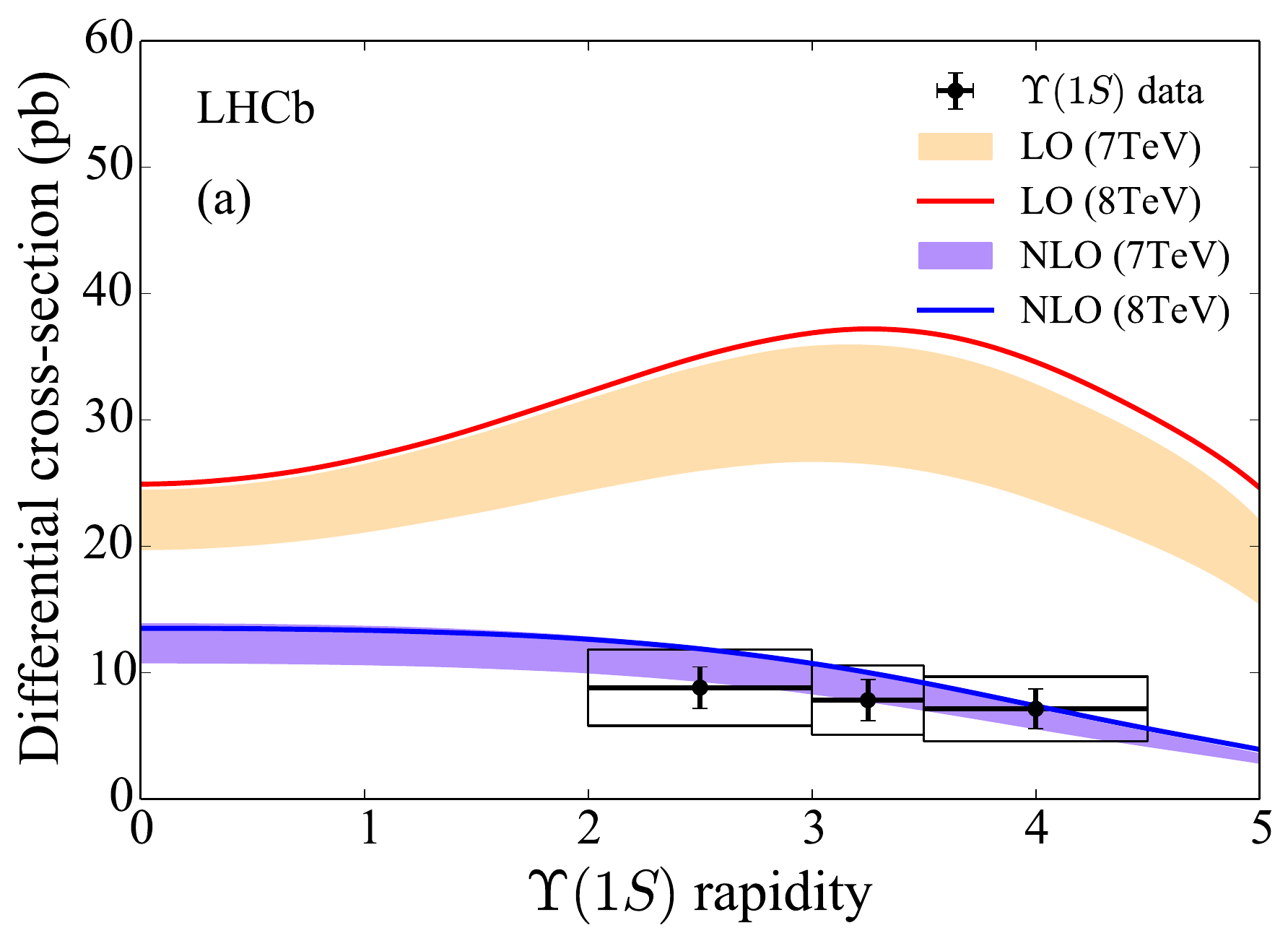}
\end{minipage}
\begin{minipage}{0.48\textwidth}\centering
\vskip-4mm\includegraphics[width=0.89\textwidth]{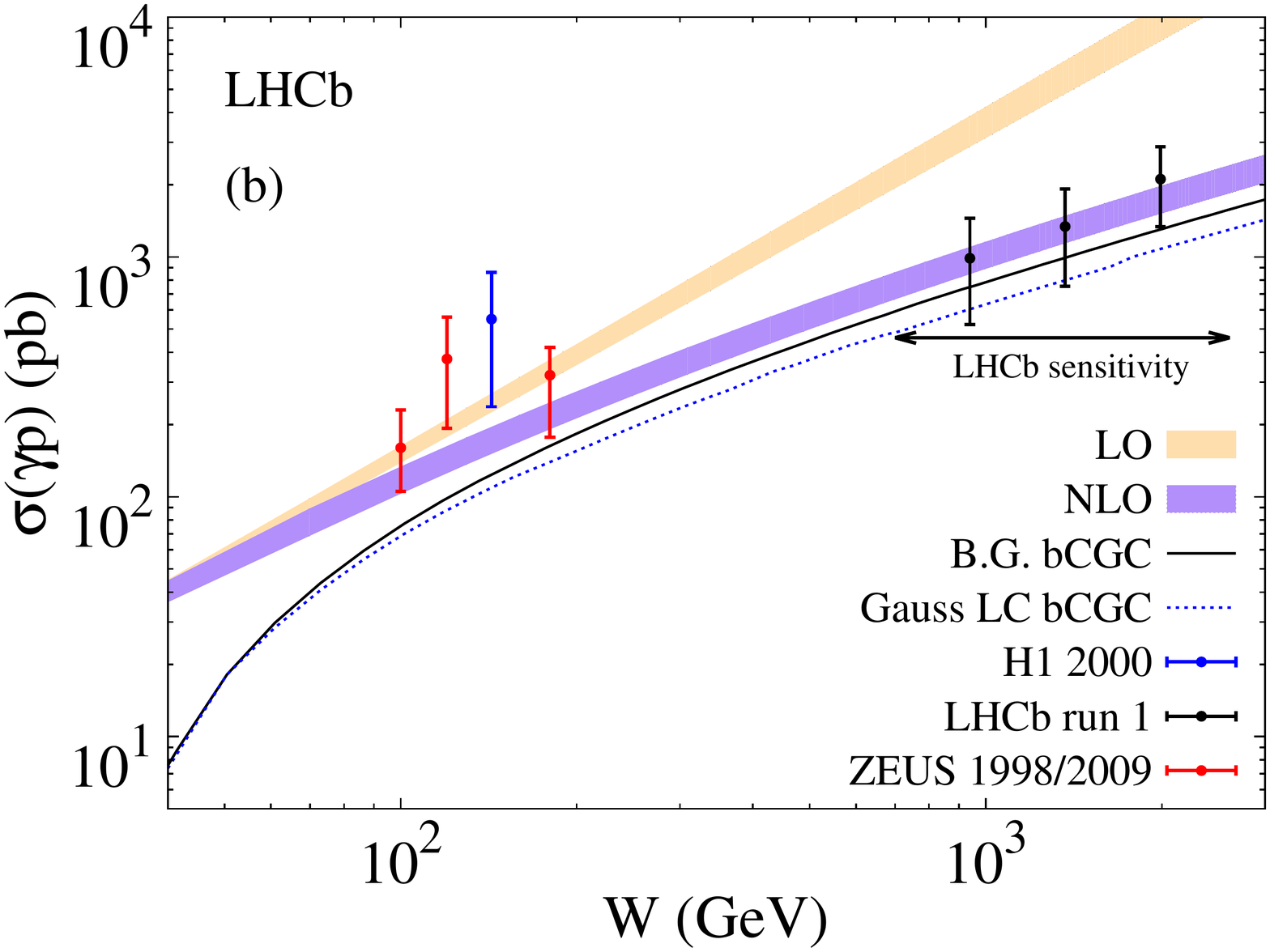}
\end{minipage}
\vskip-3mm\caption{Cross section differential in rapidity for exclusive $\Upsilon(1S)$ production (left) and exclusive $\Upsilon(1S)$ photoproduction cross section 
as a function of the photon-proton invariant mass (right). Next-to-leading-order (leading-order) JMRT calculations~\cite{jmrt} at $\sqrt{s}=7$ TeV 
are indicated by the blue (yellow) band and include uncertainties. The mean value of the next-to-leading-order (leading-order) calculations from JMRT~\cite{jmrt} at $\sqrt{s}=8$ TeV is indicated by the blue (red) line.}
\label{fig:upsi_cross}
\end{figure*}

There exist also results from exclusive $\Upsilon$ production by the LHCb collaboration, using data collected in proton-proton collisions in Run 1 at 
$\sqrt{s}=7$~TeV and $\sqrt{s}=8$~TeV, corresponding to a respective luminosity of 0.9 fb$^{-1}$ and 2.0 fb$^{-1}$~\cite{lhcb15}. The two data sets are combined in order to increase statistical precision. The data-selection procedure follows
that of the measurement for exclusive $J/\psi$, with a $p_T^2$ restricted to below $2.0$~GeV$^2/c^2$. Given the larger mass of the $\Upsilon$ meson, 
the lowest values in $x_B$ reach down to $2\times10^{-5}$. The total proton-proton production cross section for $\Upsilon(1S)$ is determined to be $9.0\pm2.1\pm1.7$ pb, 
where the first uncertainty is statistical and the second systematic, while for $\Upsilon(2S)$
 it is $1.3\pm0.8\pm0.3$ pb. For $\Upsilon(3S)$ production, an upper limit on the cross section of $3.4$ pb at the 95\% confidence level is determined. 
For the $\Upsilon(1S)$ resonance, the production cross section differential in rapidity and the photoproduction cross section as a function of $W$ are 
also extracted. They are shown in Fig.~\ref{fig:upsi_cross}. Also here, leading-order and next-to-leading order JMRT calculations are presented, and 
only the next-to-leading order calculations describe the data well. In the figure on the right, also results from the ZEUS and H1 collaborations are shown.
These are not able to discriminate between the leading-order and next-to-leading order calculations.

The LHCb collaboration also published results of exclusive production of the charmonium pairs 
$J/\psi J/\psi$, $J/\psi\psi(2S)$, $\psi(2S)\psi(2S)$, $\chi_{c0}\chi_{c0}$, $\chi_{c1}\chi_{c1}$, and $\chi_{c2}\chi_{c2}$~\cite{lhcb14p}. 
These are potentially sensitive to glueballs and tetraquarks. In the framework of describing the exclusive cross section 
in terms of standard gluon PDFs, they are sensitive to the forth power of these gluon PDFs. The measurements combine the 
data collected in proton-proton collisions 
at $\sqrt{s}=7$~TeV and $\sqrt{s}=8$~TeV. The production cross sections are measured to be $\sigma(J/\psi J/\psi)=58\pm10\pm6$~pb; $\sigma(J/\psi\psi(2S))=63^{+27}_{-18}\pm10$ pb;  
 $\sigma(\psi(2S)\psi(2S))<237$ pb; $\sigma(\chi_{c0}\chi_{c0})<69$ nb; $\sigma(\chi_{c1}\chi_{c1})<45$ pb; $\sigma(\chi_{c2}\chi_{c2})<141$ pb, 
where for the four last pairs only an upper limit is determined. These results are not corrected for proton dissociation, due to the limited 
statistical precision. Only for the production of pairs of  $J/\psi J/\psi$ it is possible to estimate the contribution from central exclusive production, 
which amounts to about 42\%, and thus to determine the cross section corrected for proton dissociation, which is $24\pm9$ pb.

\section{Summary}
Measurements of exclusive production of $J/\psi$, $\psi(2S)$, and $\Upsilon(nS)$, with $n=1,2,3$, 
have been performed by the LHCb collaboration. These measurements
are sensitive to gluon GPDs and PDFs. 
The cross sections are measured differentially in rapidity and the photoproduction cross section is extracted as a function of the photon-proton invariant mass. Comparisons to next-to-leading order JMRT calculations show good agreement with these data.  
Also cross-section measurements of pairs of charmonia have been performed. They are sensitive to the fourth power of the gluon PDFs and 
potentially to glueballs and tetraquarks. Although not discussed here, there are also preliminary measurements of Bethe-Heitler production in proton-proton 
collisions~\cite{conf_bh} and on exclusive $\chi_c$ production in proton-proton collisions~\cite{conf_bh}, which is sensitive to the exchange of 
two gluon pairs. Furthermore there are preliminary results on exclusive production of $J/\psi$ and $\psi(2S)$ in 
lead-lead collisions~\cite{lhcblead}, which give access to nuclear GPDs and PDFs, and are sensitive to shadowing.

\section{Acknowledgments}
This project has received funding from the European Union's Horizon 2020 research and innovation programme under the Marie Sklodowska-Curie grant agreement No 792684.

\end{document}